\pdfoutput=1 
\documentclass{JINST}
\usepackage{graphicx} 
\usepackage{dcolumn}  
\usepackage{longtable}
\usepackage{bm}       
\usepackage{subfigure}
\usepackage{subeqn}
\title{Multivariate side-band subtraction using probabilistic event weights}
\author{M. Williams$^a$, M. Bellis$^b$ and C. A. Meyer\\ 
Carnegie Mellon University, Pittsburgh, PA 15213, USA \\
\llap{$^a$}Current address: Imperial College London, London SW7 2AZ, UK\\
\llap{$^b$}Current address: Stanford University, Stanford, CA 94305-4060, USA  
}

\abstract{ 
A common situation in experimental physics is to have a signal which can not
be separated from a non-interfering background through the use of any 
selection criteria.
In this paper, we describe a procedure for determining, on an event-by-event 
basis, a quality factor ($Q$-factor) that a given signal candidate 
originated from 
the signal sample. This procedure generalizes the ``side-band'' 
subtraction method to higher dimensions without requiring the data to be 
divided into bins.
The $Q$-factors can then be used as event weights
in subsequent analysis procedures, allowing one to more directly access the
true spectrum of the signal.
}
\begin{document}

\section{\label{section:intro}Introduction}

A common situation in many experiments is the presence of a background
which can not be cleanly separated from the desired signal. 
If one has {\em a priori} knowledge of certain features of the signal and 
background, algorithms can be created to separate the two types of
data.
Procedures have been developed to handle many of these situations
(see, {\em e.g.},~\cite{cite:narsky,cite:cranmer,cite:vilalta}). 
Two of the more common methods for performing this type of data 
classification are neural networks and decision trees. 
In both procedures, the information known about the signal and
background is used to {\em learn} how to optimize signal-background
separation.

Consider now the case where the distributions of the signal and background are 
not known, 
and in fact, it is these distributions which we want to measure.
Of particular interest is the presence of an {\em irreducible} background,
{\em i.e.} one which can not be reduced using any selection criteria.  
An example of an irreducible background to the 
$\gamma p \rightarrow p\omega$, ${\omega\rightarrow \pi^+\pi^-\pi^0}$ reaction 
(discussed below) would be non-$\omega$ 
$\gamma p \rightarrow p\pi^+\pi^-\pi^0$ events. For any individual signal 
candidate, it is 
impossible to distinguish between these two types of data; thus, this type
of  background
can not be reduced through the use of any selection criteria.
As an example of a reducible background to this reaction, consider events where
the wrong beam photon has been associated with the tracks in the detector.  
This type of background may be reducible by examining the timing 
information of the beam and the tracks.

Perhaps the simplest method for handling an irreducible background is 
``side-band'' subtraction. 
In this procedure, distributions constructed using data from outside the 
signal region are subtracted from those built using data from inside the 
signal region in order to remove the backgrounds.
While this method can be effective in some situations, implementing it can
become problematic if the kinematics of the background region are different
than those of the signal region or if the problem is sufficiently 
multi-dimensional such that binning the data is severely limited by statistical
uncertainties.

In this paper, we describe a procedure for generalizing the one-dimensional
side-band subtraction method to higher dimensions without having to bin the 
data. Our method involves using {\em nearest neighbor} events
to assign each signal candidate (henceforth referred to as an ``event'')
in a data sample a quality factor ($Q$-factor) which gives the probability 
that it 
originates from the signal sample. The data are assumed to be described by
a set of coordinates which can be masses, angles, energies, {\em etc.}
The distributions of the signal and background must be known 
(possibly with unknown parameters) in a subset of the coordinates, referred to
as the {\em reference} coordinates.  No {\em a priori} information concerning
the signal or background distributions in any other coordinate is required.
Thus, parametrizations of the signal and background are not necessary in any of
the non-reference coordinates. The unknown parameters in the signal and 
background reference-coordinate probability density functions (PDFs) are 
determined {\em locally} in the non-reference coordinates, 
{\em i.e.} they are allowed to vary according to the non-reference 
coordinates. No information, not even parametrizations, about how these 
parameters depend on the non-reference coordinates is required 
(see Section~\ref{section:method} for more details).  
This is an extremely useful property of this method and one that makes its
use in certain analyses advantageous over many other methods.
Once the $Q$-factors are obtained, the data in the side bands can be discarded
(this may be desirable in some analyses).

The $Q$-factors can be used as event weights
in subsequent analysis procedures to gain access to the signal distribution.
For example, the $Q$-factors can be used in an event-based unbinned 
maximum likelihood fit performed on the data to extract physical observables.
By computing the weighted sum of log likelihoods, the background subtraction is
carried out automatically in the fit without ever having to resort to dividing 
the data up into bins.
Eliminating the need to bin the data is highly desirable for the case of 
multi-dimensional problems. 

In this paper we assume that no correlation exists between the reference 
coordinates and the remaining set of coordinates. 
In principle, this limitation could be overcome in some analyses
(see Section~\ref{section:method}). 
It is also assumed that 
there is no quantum mechanical interference between the signal and background.
One final assumption is that the signal and background distributions do not 
vary rapidly in the non-reference coordinates relative
to the {\em correlation distances}, {\em i.e.} the diameters of the 
hyper-spheres used to collect the nearest neighbor events. 
A similar constraint exists in binned analyses. If a 
distribution varies rapidly relative to the bin width, then some of the finer
structure present in the distribution will be lost.  
This same ``averaging'' effect can also occur in our method if the signal 
distribution possesses rapid variations in the non-reference coordinates.

The idea of using nearest neighbor events as a means of data classification is 
not
new (see, {\em e.g.},~\cite{cite:hastie}).  Other methods also exist which 
exploit 
the concept of reference coordinates to separate out contributions from 
different sources to a data set.  A good example is the 
${}_{s}\mathcal{P}lot$ technique~\cite{cite:splot}.  The method presented in
this paper is unique in 
that it combines these ideas to carry out an unbinned side-band subtraction 
which results in each event obtaining an event weight. 
The only information required as input are parametrizations of the signal and
background in terms of the reference coordinates.  No {\em a priori} 
knowledge about the signal or background in the remaining coordinates is 
necessary.  Another advantage of this method is that it permits the presence 
of unknown parameters in the signal and background reference-coordinate PDFs.
These parameters are determined locally in the non-reference coordinates; 
thus, they are allowed to vary in
these coordinates and no parametrization of these variations is required.
The event weights obtained can be used in 
an unbinned fit to extract physical observables from the data. 

As an example, in Section~\ref{section:example} we will consider the reaction 
${\gamma p \rightarrow p \omega}$, $\omega \rightarrow \pi^+\pi^-\pi^0$.
There are, of course, production mechanisms other than  
${\gamma p \rightarrow p \omega}$ which produce the same final state and 
no selection criteria exists which can separate out the
${\gamma p \rightarrow p \omega}$ events (the background is irreducible). 
The only knowledge we have about the background is that it can be parametrized
by a polynomial (whose parameters are unknown) in the three-pion mass.  The 
distribution of the background in all other variables is completely unknown.
The goal of our model analysis will be to measure the $\omega$ polarization
observables known as {\em spin density matrix elements},   
which can be extracted from the distribution
of the pions in the $\omega$ rest frame.  Ideally, we would like to avoid 
binning the data and to extract these observables from an event-based maximum
likelihood fit.  The method presented in this paper will allow us to perform
such an analysis.

\section{\label{section:method}Quality Factor Determination}

Consider a data set composed of $n$ total events, 
each of which is described 
by $m$ coordinates $\vec{\xi}$ (${m \geq 2}$).
Furthermore, the data set consists of $n_s$ events which are signal
and $n_b$ events which are background. 
Both the signal and background distributions 
are functions of the coordinates, $S(\vec{\xi})$ and $B(\vec{\xi})$, 
respectively. 
For this procedure, we need to know the functional dependence (unknown 
parameters are permitted) of
the signal and background distributions in terms of (at least) one of the 
coordinates.
We will refer to this coordinate as the reference coordinate and label it
$\xi_r$.  It is trivial to extend this method to consider any number of 
reference coordinates if necessary.  

As an example, consider the case where the reference coordinate is a
mass. The functional dependence of the signal, in terms of $\xi_r$, might be 
given by a Gaussian or Breit-Wigner distribution. 
The background may be well represented
by a polynomial. In both cases, there could be unknown parameters
({\em e.g.} the width of the Gaussian); these are permitted when using this
procedure. These unknown parameters can vary in terms of the non-reference
coordinates.  For example, the width of a Gaussian describing the mass 
of a composite 
particle may depend on the lab angles of its decay products. The unknown 
parameters in the PDFs are determined locally in the non-reference coordinates;
thus, these types of variations are handled automatically by the method.
No other {\em a priori} information is required concerning the 
dependence of $S(\vec{\xi})$ or $B(\vec{\xi})$ on any of the other 
coordinates.

The aim of this procedure is to assign each event a quality factor, 
or 
$Q$-factor, which gives the probability 
that it originates from the signal sample.
We first need to define a metric for the space spanned by $\vec{\xi}$
(excluding $\xi_r$). A reasonable choice is to use 
$\delta_{kl}/\sigma^2_k$ where $\sigma_k$ is the root mean square (RMS) of the
$k^{th}$ variable in the appropriate phase-space distribution 
(see Section~\ref{section:example} for some examples). 
This gives equal weight to each variable. 
Some care should be taken when choosing a metric and the choice presented 
here may not be the optimal one for all analyses.  Discussion on this topic
can be found at the end of this section.
Using this metric, the distance between any two events, $d_{ij}$, 
is given as
\begin{equation}
  \label{eq:dist}
  d^2_{ij} = \sum\limits_{k \neq r} \left[ \frac{\xi^i_k - \xi^j_k}
    {\sigma_k} 
    \right]^2,
\end{equation}
where the sum is over all coordinates except $\xi_r$. This is known as the
{\em normalized Euclidean distance}.

For each event, we compute the distance to all other events 
in the data
set, and retain the $n_c$ nearest neighbor events, 
including the event itself, 
according to Eq.~(\ref{eq:dist}). 
The value of $n_c$, which varies depending on the
analysis, is discussed below and in Section~\ref{section:example:nc}. 
It is worth noting that the limit $n_c \rightarrow n$ is equivalent to 
performing a global side-band subtraction ({\em i.e.} determining the total
number of signal events in the data set).
The $n_c$ events are then fit using the 
unbinned maximum likelihood method to obtain estimators for the parameters, 
$\vec{\alpha}$, in the PDF
\begin{equation}
  F(\xi_r,\vec{\alpha}) = 
  \frac{F_s(\xi_r,\vec{\alpha}) + F_b(\xi_r,\vec{\alpha})}
  {\int\left[F_s(\xi_r,\vec{\alpha}) + F_b(\xi_r,\vec{\alpha})\right]d\xi_r},
\end{equation}
where $F_s$ and $F_b$ describe the functional dependence on the reference
coordinate, $\xi_r$, of the signal and background, respectively. 
These distributions are normalized such that for any given set of 
estimators, $\hat{\alpha}$,
\begin{equation}
  \int F_s(\xi_r,\hat{\alpha})d\xi_r = n_{sig}, \hspace{0.01\textwidth}
    \int F_b(\xi_r,\hat{\alpha})d\xi_r = n_{bkgd},
\end{equation}
where $n_{sig}(n_{bkgd})$ is the total amount of signal (background) 
extracted from the $n_c$ nearest neighbor event sample. 
We now return to the assumption in Section~\ref{section:intro} that the 
reference and non-reference coordinates are uncorrelated.  
The motivation for this assumption should now be clear.  If there is a strong
correlation between these two types of coordinates, it could lead to 
distortions in the reference-coordinate distributions constructed out of the
$n_c$ nearest neighbor events resulting in a bias in the extracted $Q$-factors.
In principle, this could be overcome if these distortions can properly be 
accounted for in the PDFs.  While this possibility is intriguing, its
validity is not tested in this paper as it is likely to be very problem
specific.  

The $Q$-factor for each event is then calculated as
\begin{equation}
  \label{eq:q-factor}
  Q_i = \frac{F_s(\xi^i_r,\hat{\alpha}_i)}
  {F_s(\xi^i_r,\hat{\alpha}_i) + F_b(\xi^i_r,\hat{\alpha}_i)},
\end{equation}
where $\xi_r^i$ is the value of the event's reference coordinate and 
$\hat{\alpha}_i$ are the estimators for the parameters obtained from the
event's fit. Since a separate fit is run to determine the $\hat{\alpha}_i$
values for each event using its $n_c$ nearest neighbors, the estimators 
obtained are the local values for the hyper-sphere around the $i^{th}$ event.
Thus, if they vary according to the non-reference coordinates, these 
variations will be automatically accounted for provided they do not vary 
rapidly relative to the correlation distance (one of our stated assumptions
in Section~\ref{section:intro}).
We note here that a similar looking construct involving likelihood
ratios (also denoted by $Q$) is used in high-energy physics to determine 
discovery significance (see, {\em e.g.},~\cite{cite:read}). The ratio in
Eq.~(\ref{eq:q-factor}) is built from terms using estimators obtained from the
same fit; thus, it is quite different from its likelihood ratio namesake.

If one wants to bin the data, the signal yield in a bin is obtained as
\begin{equation}
  \label{eq:sig-yield}
  \mathcal{Y} = \sum\limits_i^{n_{bin}} Q_i,
\end{equation}
where $n_{bin}$ is the number of events in the bin. For example, to
construct a histogram (of any dimension) of the signal, one would simply 
weight each event's contribution by its $Q$-factor. 

The metric presented in this section is sufficient for many high-energy 
physics analyses;
however, there are some cases where it may not be the optimal choice.  
One of our stated assumptions in Section~\ref{section:intro} was that the 
distributions should not vary rapidly relative to the correlation 
distance (which is determined by the metric and statistics).
This could result in the loss of some of the finer structure in the signal. 
If the distributions are expected or observed to vary more rapidly in 
some subset of the coordinates, then it may be necessary to use a metric which
can handle this situation.
Thus, for some analyses it may be necessary to weight the coordinates 
{\em unequally}
in the metric or to construct a totally different metric all together. 
We can again compare this situation to a similar one in binned analyses.  
Consider a two-dimensional analysis where the physics is known to vary 
rapidly in coordinate $x$ and slowly in coordinate $y$.  
In this case, one would simply bin the data finer in $x$ than in $y$ prior to 
analyzing it.  This asymmetric binning is analogous to using 
unequal weights in the metric in our procedure.

From this discussion one can clearly see that the choice of metric can effect
the results, {\em i.e.} the metric effects how the $n_c$ nearest
neighbor events are obtained.
This is analogous to the fact that in a binned analysis choosing a different
binning scheme can effect the results.  There is no way to quantitatively 
determine what is the optimal binning. This is determined in an {\em ad-hoc}
manner in each analysis.  Typically, one wants the bin population to be high
to reduce the relative statistical uncertainty. This dictates choosing a 
large bin width. If, however, the bin width is large relative to the variations
in the distributions, then the finer structure will be lost.  
Thus, there are two competing factors which should be considered when choosing 
a binning.  In practice, any {\em reasonable} choice should not effect the 
observables extracted from the data.  

This same {\em ad-hoc} approach is necessary in our method when choosing a 
metric and a value for $n_c$.  
The statistical uncertainties on the $Q$-factors increase as $n_c$ decreases. 
Clearly, we would like to keep this uncertainty as small as possible. 
This would dictate choosing a large value for $n_c$.  
If, however, $n_c$ is large relative to $n$, then the method is averaging over 
large fractions of phase space. 
This can result in losses of the finer structure in the distributions.
Thus, the factors to consider when choosing a value for $n_c$ are the same as
when choosing a bin width in a binned analysis.

It cannot be stressed enough that the validity of the method presented in this
paper depends on how rapidly the coordinates vary relative to the correlation 
distance. The correlation distance is determined by the signal and background
PDFs, the metric and the value of $n_c$.  Unfortunately, it is not possible
to provide a general prescription for either the metric or the value of $n_c$ 
that will guarantee the validity of the method for every analysis.
Instead, the choice of metric and $n_c$ must be studied in the data and also 
in Monte Carlo data (see Section~\ref{section:example:nc}).
As in a binned analysis, any {\em reasonable} choice should not effect the 
extracted observables. 
We conclude this section by noting that for cases with very high 
dimensionality, it may be necessary 
to work in ``Gaussianized'' variables~\cite{cite:chen} instead of the measured 
quantities.

\section{\label{section:errors}Error Estimation}

It is also important to extract the uncertainties on the individual 
$Q$-factors so that we can obtain error estimates on measurable 
quantities.
The full covariance matrix obtained from each event's fit, 
$C_{\alpha}$, can be used to calculate the uncertainty in $Q$ as
\begin{equation}
  \label{eq:sigma-q}
  \sigma^2_Q = \sum \limits_{ij} \frac{\partial Q}{\partial \alpha_i}
  (C_{\alpha}^{-1})_{ij} \frac{\partial Q}{\partial \alpha_j}.
\end{equation}
When using these values to obtain errors on the signal yield in any 
bin, we must consider the fact that the nature of our procedure leads to 
highly-correlated results for each event and its $n_c$ nearest 
neighbors.
The uncertainty on the signal yield due to signal-background separation 
in a bin is properly given by
\begin{equation}
  \label{eq:q-err-prop}
  \sigma_{\mathcal{Y}}^2 = \sum\limits_{i,j} \sigma_Q^i \rho_{ij} \sigma_Q^j,
\end{equation}
where the sums $(i,j)$ are over the events in the bin and $\rho_{ij}$ is the
correlation factor between events $i$ and $j$.  
This factor is equal to the 
fraction of shared nearest neighbor events 
used in calculating the $Q$-factors for these events. 

Keeping track of these correlations can be a bit cumbersome (though, 
it is possible).  An overestimate of the true uncertainty inherent in the 
procedure can be obtained by assuming 100\% correlation as follows:
\begin{equation}
  \label{eq:q-err}
  \sigma_{\mathcal{Y}} = \sum\limits_{i}^{n_{bin}} \sigma_{Qi},
\end{equation}
How much this approximation overestimates the errors depends on the population
of the bin; however, it is often a decent estimate due to the similar
constraints which factor into the choices of bin size and the value of 
$n_c$ (see Section~\ref{section:example:nc}).

In addition to the uncertainties associated with the fits, there will also be
a purely statistical uncertainty associated with the signal yield in each bin, 
given by Poisson statistics as follows:
\begin{equation}
  \label{eq:q-err-stat}
  \sigma^2_{\mathcal{Y}stat} = \sum\limits_{i}^{n_{bin}} Q_i^2.
\end{equation}
The total uncertainty on the signal yield in any bin is then obtained by adding
the fit errors, calculated using Eq.~(\ref{eq:q-err}), in quadrature with the
statistical errors obtained from Eq.~(\ref{eq:q-err-stat}). 

The variation of the correlation distance as a function of the non-reference
coordinates makes proper handling of the errors an intricate task.  
The method for determining the uncertainty on the signal yield in any bin
presented in this section is rigorous; however, the impact of the correlations
between the $Q$-factors and their uncertainties on physical observables 
extracted from the data is not as straight-forward.  For example, if the 
data is binned, then correlations will also be present between the bins 
(although, they should be small). 
It is strongly recommended that a Monte Carlo study be performed to 
ensure that the uncertainties on physical observables extracted from the data
are being handled correctly.  

\section{\label{section:obs}Extracting Observables: Event-Based Fitting using $Q$-Factors}

One of the primary motivations behind the development of this method was to 
make it 
possible to extract physical observables from multi-dimensional distributions
without having to resort to binning the data.
The $Q$-factors obtained for each event above can be used in conjunction with
the unbinned maximum likelihood method to avoid this difficulty.

If we could cleanly separate out the signal from the background, then 
the likelihood function would be defined as
\begin{equation}
  \mathcal{L} = \prod\limits_i^{n_s} W(\vec{\xi}),
\end{equation}
where $W$ is some PDF (with unknown parameters) which the data is to be fit
to.
We could then obtain estimators for the unknown parameters in $W$ by 
minimizing
\begin{equation}
  \label{eq:log-L}
  -\ln{\mathcal{L}} = -\sum\limits_i^{n_s} \ln{W(\vec{\xi}_i)}.
\end{equation}
For cases where it is not possible to separate the signal and background 
samples, we can use the $Q$-factors to achieve the same effect
by rewriting Eq.~(\ref{eq:log-L}) as
\begin{equation}
  \label{eq:log-L-Q-def}
  -\ln{\mathcal{L}} = -\sum\limits_i^{n} Q_i \ln{W(\vec{\xi}_i)},
\end{equation}
where the sum is now over all events (which contains both signal
and background).
Thus, the $Q$-factors are used to weight each event's contribution to the 
likelihood.

\section{\label{section:example}Example Application}

As an example, we will consider the reaction ${\gamma p \rightarrow p \omega}$
in a single $(s,t)$ bin, {\em i.e.} a single center-of-mass energy and 
production angle bin (extending the example to avoid binning in
production angle, or $t$, is discussed below). The $\omega$ decays to
$\pi^+\pi^-\pi^0$ about 90\% of the time; thus, we will assume we have a 
detector which has reconstructed ${\gamma p \rightarrow p \pi^+\pi^-\pi^0}$
events. Of course, there are production mechanisms other than  
${\gamma p \rightarrow p \omega}$ which can produce this final state and there
is no selection criteria 
which can separate out events that originated from
${\gamma p \rightarrow p \omega}$. Below we will construct a toy-model of this
situation by generating Monte Carlo events for both signal, {\em i.e.}
$\omega$ events, and background, {\em i.e.} non-$\omega$ $\pi^+\pi^-\pi^0$
events (10,000 events were generated for each).
The goal of our model analysis is to extract the $\omega$ polarization
observables known as the spin density matrix elements, denoted by
$\rho^0_{MM'}$ (discussed below).  We note here that in this example we will
assume that the signal and background do not interfere (and generate the Monte
Carlo data accordingly).  In real data, for this reaction 
a small amount of interference would be expected 
due to the 8.44~MeV natural width of the $\omega$.

\begin{figure}
  \centering
  \includegraphics[width=0.5\textwidth]{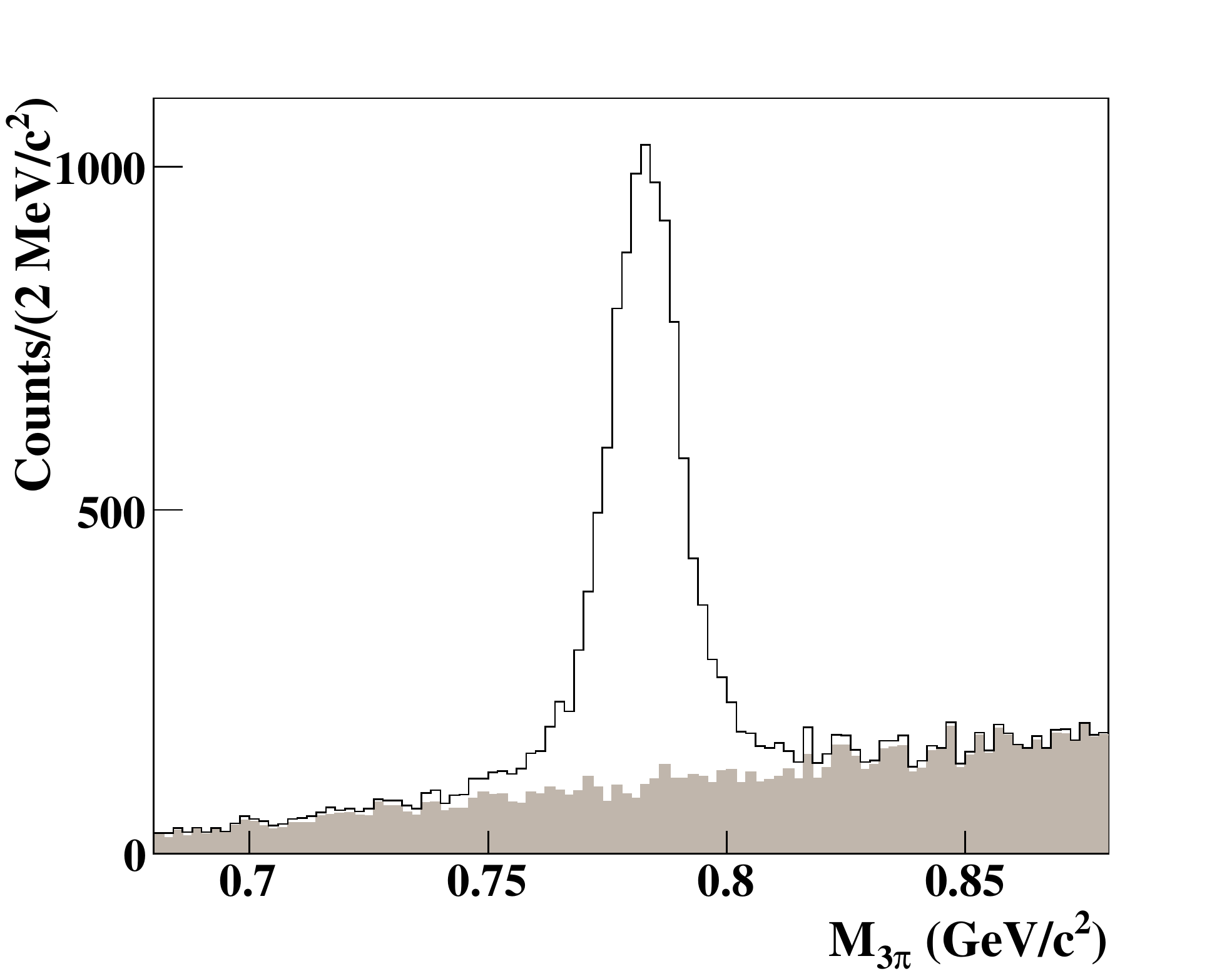}
  \caption[]{\label{fig:m3pi-gen}
    Mass of the $\pi^+\pi^-\pi^0$ system (GeV/c$^2$) for all generated events
    (unshaded) and for only the background (shaded).
  }
\end{figure}

In terms of the mass of the $\pi^+\pi^-\pi^0$ system, $m_{3\pi}$, the $\omega$
events were generated according to 3-body phase space weighted by a 
Voigtian (a convolution of a Breit-Wigner and a Gaussian, see 
Eq.~(\ref{eq:voigt})) to account for both the natural width of the $\omega$ and
detector resolution. For this example, we chose to use $\sigma = 5$~MeV/c$^2$
for the detector resolution (see Fig.~\ref{fig:m3pi-gen}). 
The goal of our analysis is to extract the three measurable elements of the
spin density matrix (for the case where neither the beam nor target are 
polarized) traditionally chosen to be $\rho^0_{00}$, $\rho^0_{1-1}$ and 
$Re\rho^0_{10}$. These can be accessed by examining the 
distribution of the decay products ($\pi^+\pi^-\pi^0$) of the $\omega$ in its
rest frame.

\begin{figure*}
  \begin{center}
  \includegraphics[width=1.0\textwidth]{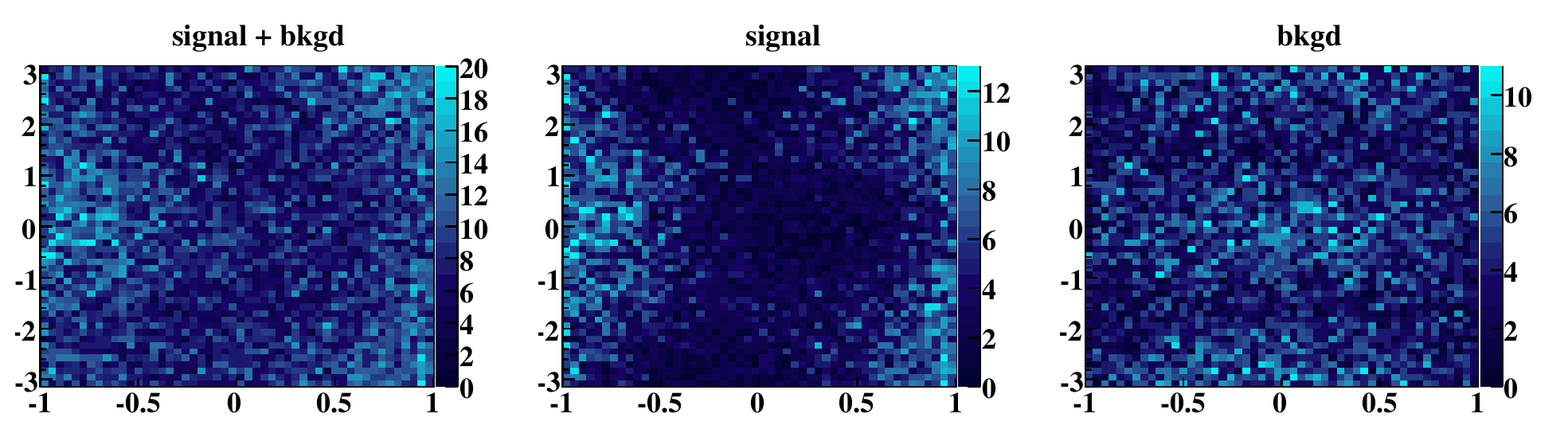}
  \caption[]{\label{fig:decay-angles-gen}
    (Color Online)
    $\phi$ (radians) vs $\cos{\theta}$: Generated decay angular distributions
    for all events (left), only signal events (middle) and only background
    events (right).
  }
  \end{center}
\end{figure*}

For this example, we chose to work in the helicity system which defines the
$z$ axis as the direction of the $\omega$ in the overall center-of-mass 
frame, the $y$ axis as the normal to the production plane and the $x$ axis is
simply given by ${\hat{x} = \hat{y} \times \hat{z}}$.
The decay angles $\theta,\phi$ are the polar and azimuthal angles of the normal
to the decay plane in the $\omega$ rest frame, {\em i.e.} the angles of the
vector ${\left(\vec{p}_{\pi^+} \times \vec{p}_{\pi^-}\right)}$.
The decay angular distribution of the $\omega$ in its rest frame is then 
given by~\cite{cite:schilling-1970}
\begin{eqnarray}
  \label{eq:schil}
  W(\theta,\phi) 
  = \frac{3}{4\pi} \left(\frac{1}{2}(1 - \rho^0_{00})
  + \frac{1}{2}(3\rho^0_{00} - 1)\cos^2{\theta}
  \right.\hspace{0.3\textwidth}
  \nonumber\\
  \hspace{0.3\textwidth}
  - \left.
  \rho^0_{1-1}\sin^2{\theta}\cos{2\phi}
  - \sqrt{2}Re\rho^0_{10}\sin{2\theta}\cos{\phi}\right),
\end{eqnarray}
which follows directly from the fact that the $\omega$ is a vector particle;
it has spin-parity $J^P = 1^-$. 
We chose to use the following $\rho^0_{MM'}$ values for this example:
\begin{subequations}
  \label{eq:rho-gen}
  \begin{equation}
  \rho^0_{00} = 0.65 
  \end{equation}
  \begin{equation}  
  \rho^0_{1-1} = 0.05 
  \end{equation}
  \begin{equation}
    Re\rho^0_{10} = 0.10
  \end{equation}
\end{subequations}
The resulting generated decay distribution is shown in 
Fig.~\ref{fig:decay-angles-gen}.

For the background, we chose to generate it according to 3-body phase space 
weighted by a linear function in $m_{3\pi}$ and 
\begin{equation}
  \label{eq:bkgd-pdf}
  W(\theta,\phi) = \frac{1}{6\pi}\left(1 + |\sin{\theta}\cos{\phi}|\right)
\end{equation}
in the decay angles. When separating out the signal below, we assume we have
no knowledge of Eq.~\ref{eq:bkgd-pdf} (this would typically be the case).
Thus, the separation of signal and background will be carried out using only
the knowledge that the background can be parametrized as a polynomial (with
unknown parameters) in $m_{3\pi}$.  No information about the distributions of 
the background in any other variables will be used.
Figure~\ref{fig:m3pi-gen} shows the $\pi^+\pi^-\pi^0$ mass
spectrum for all generated events and for just the background. The generated
decay angular distributions for all events, along with only the signal and
background are shown in Fig.~\ref{fig:decay-angles-gen}. There is clearly no
selection criteria which can separate out the signal.

\subsection{Applying the Procedure}
\label{section:example:apply}

To obtain the $Q$-factors, we first need to identify the relevant coordinates,
{\em i.e.} the kinematic variables in which we need to separate signal 
from background. The $\pi^+\pi^-\pi^0$ mass will be used as the reference 
coordinate, $\xi_r \equiv m_{3\pi}$.
The stated goal of our analysis is to extract the 
$\rho^0_{MM'}$ elements. We will do this using Eq.~(\ref{eq:schil}); thus, only
the angles $\theta,\phi$ are relevant. Other decay variables, such as the
distance from the edge of the $\pi^+\pi^-\pi^0$ Dalitz plot, are not relevant
to this analysis --- though, they would be in other analyses 
(see Section~\ref{section:example:full-pwa}). 

Using the notation of
Section~\ref{section:method}, $\vec{\xi} = (m_{3\pi},\cos{\theta},\phi)$.
The RMS's of the relevant kinematic variables are
\begin{subequations}
\begin{equation}
\sigma^2_{\phi} = \int_{-\pi}^{\pi}\phi^2d\phi = 2\pi^3/3 \\
\end{equation}
\begin{equation}
\sigma^2_{\cos{\theta}} = \int_{-1}^{1}\cos^2{\theta}d\cos{\theta} =
 2/3.
\end{equation}
\end{subequations}
The distance between any two points, $d_{ij}$, is then given by
\begin{equation}
  d^2_{ij} = \frac{3}{2}\left[(\cos{\theta_i} 
      - \cos{\theta_j})^2 
    + \frac{(\phi_i - \phi_j)^2}{\pi^3}\right].
\end{equation}
The functional dependence of the signal and background on the reference 
coordinate, $m_{3\pi}$, are
\begin{subequations}
\begin{equation}
  F_s(m_{3\pi},\vec{\alpha}) = 
  s\cdot V(m_{3\pi},m_{\omega},\Gamma_{\omega},\sigma)
\end{equation}
\begin{equation}
  F_b(m_{3\pi},\vec{\alpha}) = b_1 m_{3\pi} + b_0,  
\end{equation}
\end{subequations}
where $m_{\omega} = 782.56$~MeV/c$^2$, $\Gamma_{\omega} = 8.44$~MeV, 
${\sigma=5}$~MeV is the simulated detector resolution, 
${\vec{\alpha} = (s,b_1,b_0)}$ are unknown parameters and 
\begin{equation}
  \label{eq:voigt}
  V(m_{3\pi},m_{\omega},\Gamma_{\omega},\sigma)  =
  \frac{1}{\sqrt{2\pi}\sigma} Re \left[
    w\left(\frac{1}{2\sqrt{\sigma}}(m_{3\pi} - m_{\omega}) 
    + i\frac{\Gamma_{\omega}}{2\sqrt{2}\sigma}
    \right)\right],
\end{equation}
is the convolution of a Gaussian and non-relativistic Breit-Wigner
known as a Voigtian ($w(z)$ is the complex error function). 

As stated above, the number of nearest neighbor events 
required depends on the
analysis. Specifically, it depends on how many unknown parameters there are, 
along with the functional forms of $F_s$ and $F_b$. For this relatively simple
case, the value $n_c = 100$ works well (see Section~\ref{section:example:nc}
for discussion on the value of $n_c$).
For each simulated event, we then
find the $n_c$ closest events 
(containing both signal and background events) and 
perform an unbinned maximum likelihood fit, using the CERNLIB package 
MINUIT~\cite{cite:minuit}, to determine the estimators $\hat{\alpha}$.  
The $Q$-factors are then calculated from Eq.~(\ref{eq:q-factor})
and the uncertainties are straightforward to calculate following
Section~\ref{section:errors}.

\begin{figure}
  \centering
  \includegraphics[width=0.5\textwidth]{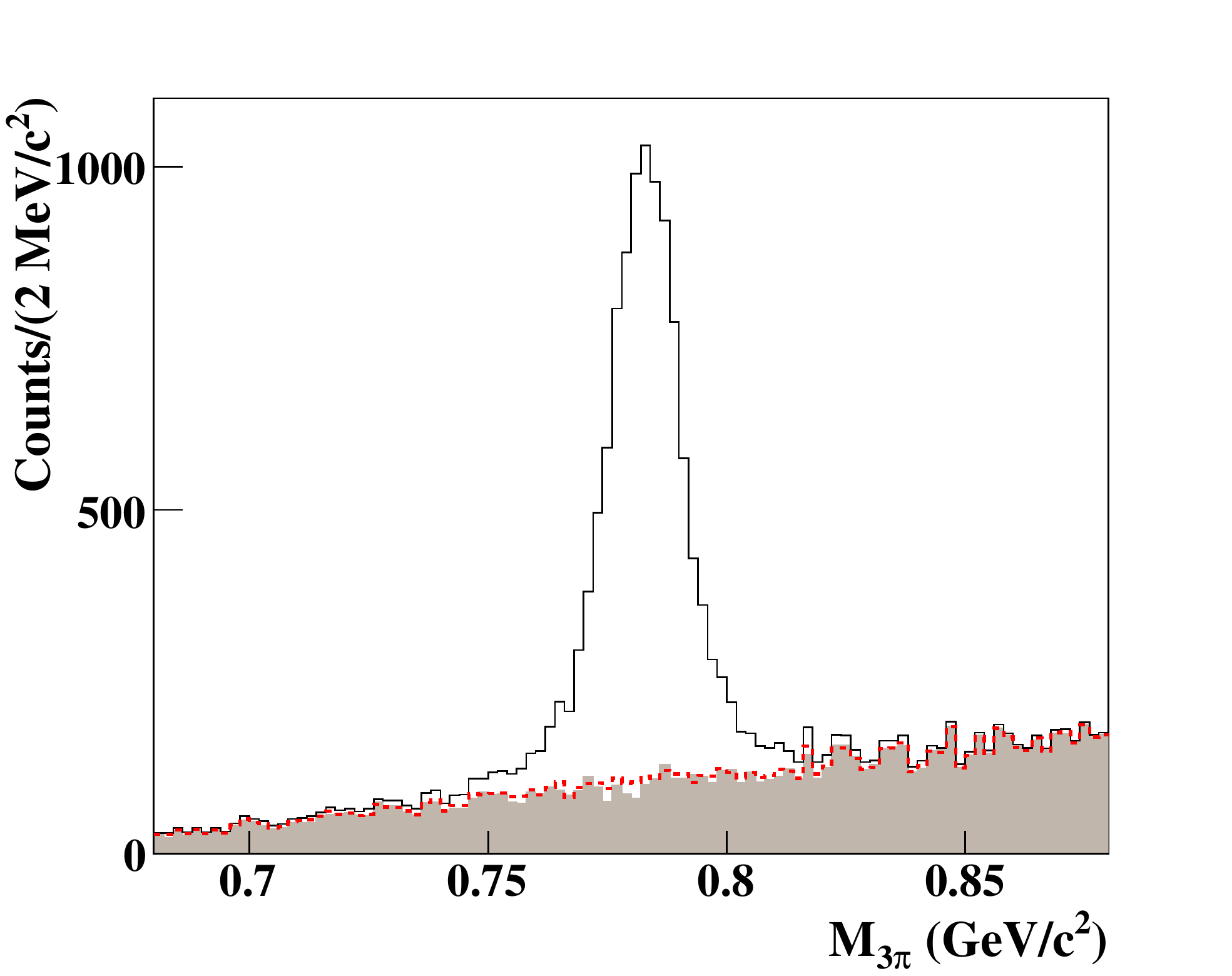}
  \caption[]{\label{fig:m3pi-fit}
    (Color Online)
    Mass of the $\pi^+\pi^-\pi^0$ system (GeV/c$^2$) for all generated events
    (unshaded), only generated background events (shaded) and all generated
    events weighted by $1-Q$ (dashed-red).
  }
\end{figure}

\begin{figure*}
  \begin{center}
  \includegraphics[width=1.0\textwidth]{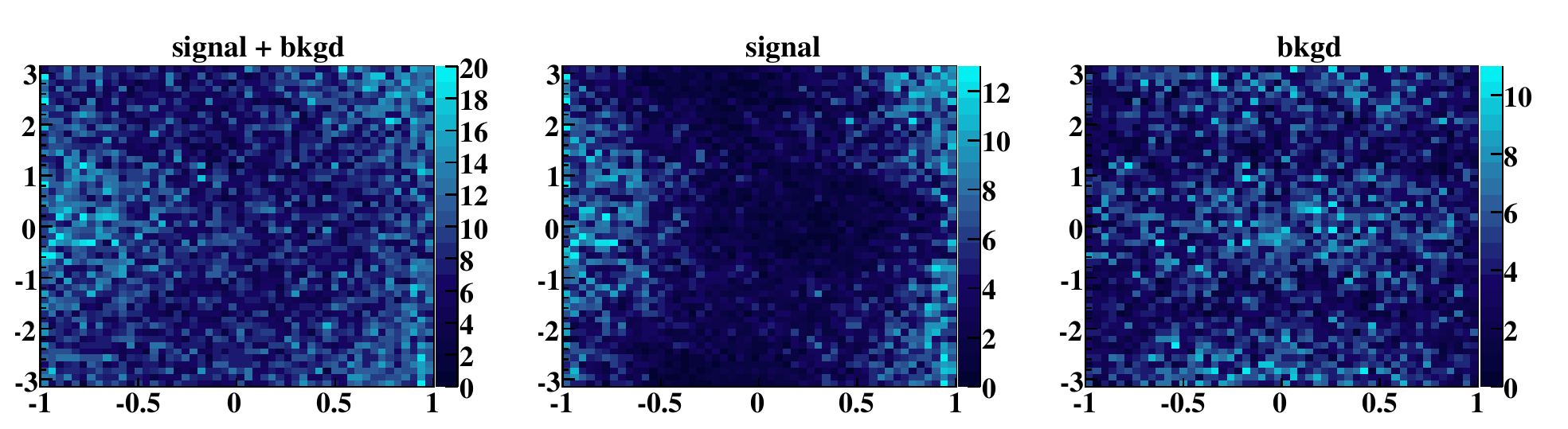}
  \caption[]{\label{fig:decay-angles-fit}
    (Color Online)
    $\phi$ (radians) vs $\cos{\theta}$: Extracted decay angular distributions
    for all events (left), for events weighted by $Q$, 
    signal (middle), and for events weighted by $1-Q$, background (right).
  }
  \end{center}
\end{figure*}

\begin{figure*}
  \centering
  \subfigure[]{
    \includegraphics[width=0.49\textwidth]{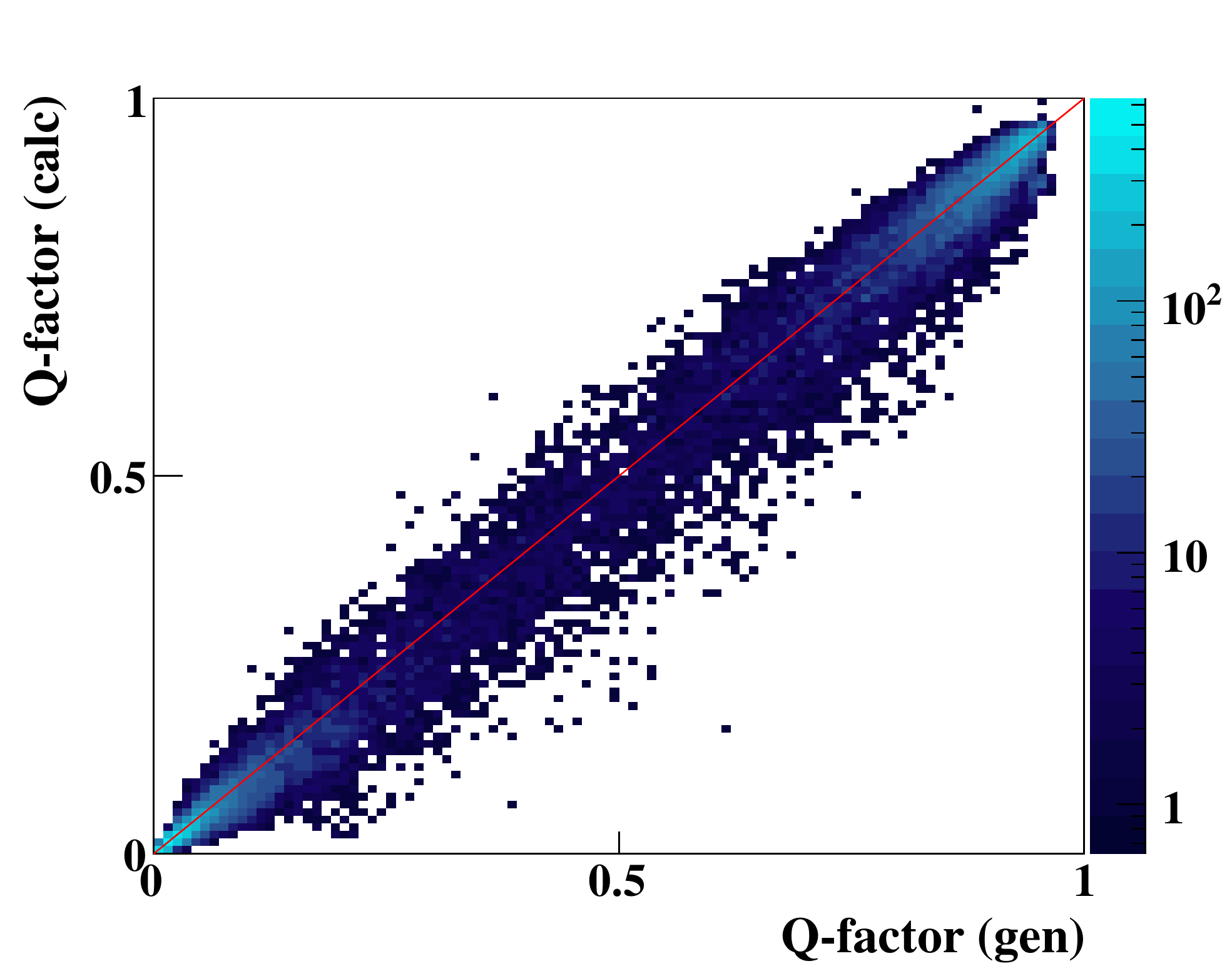}
  }
  \hspace{-0.02\textwidth}
  \subfigure[]{
    \includegraphics[width=0.49\textwidth]{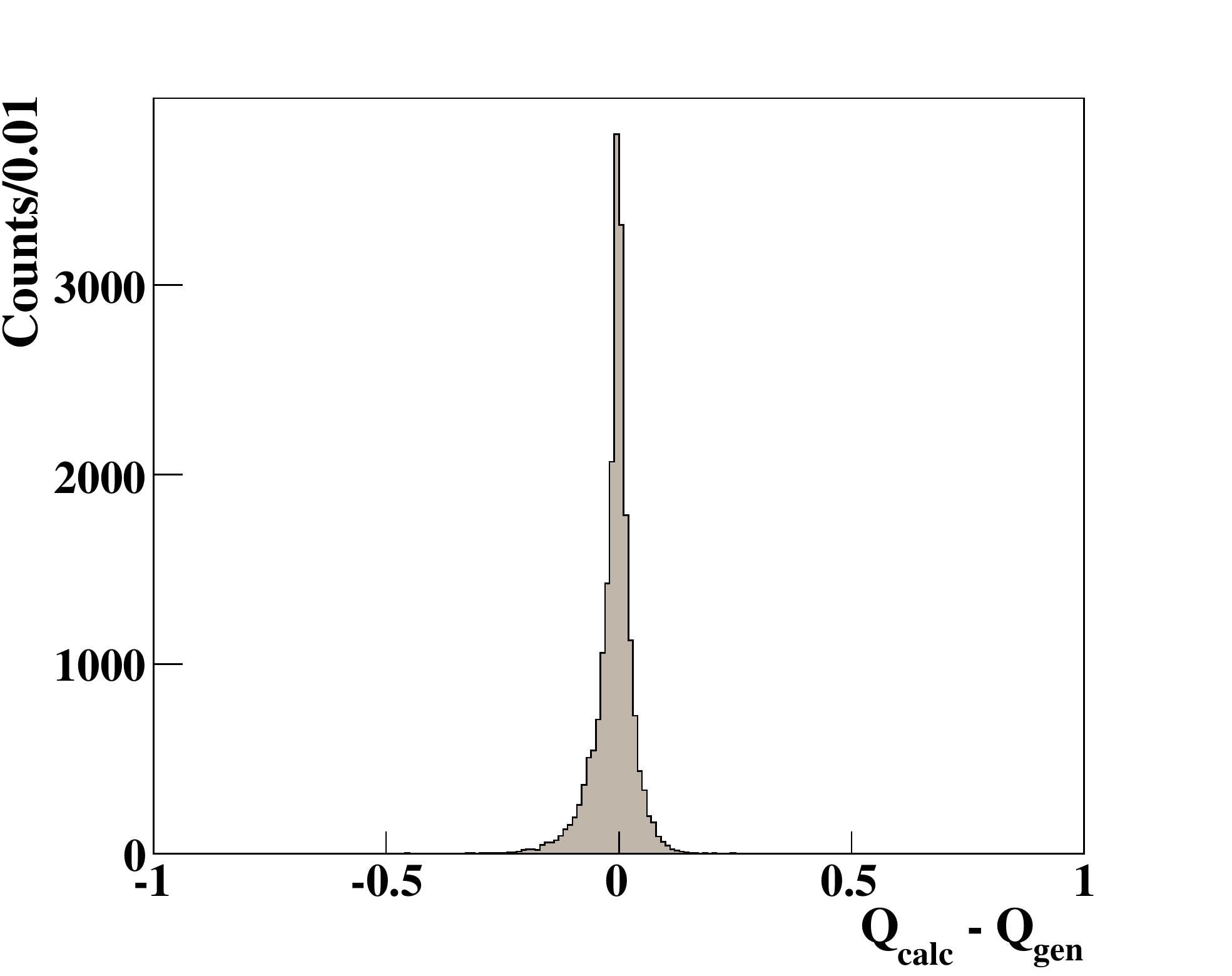}
  }
  \caption[]{\label{fig:q-compare}
    (Color Online)
    (a) Calculated $Q$-factors vs generated $Q$-factors. The red line indicates
    where ${Q_{calc} = Q_{gen}}$.
    (b)~${Q_{calc} - Q_{gen}}$: The difference between the generated and 
    calculated $Q$-factors.
  }
\end{figure*}

Figure~\ref{fig:m3pi-fit} shows the comparison of the extracted and generated 
background $m_{3\pi}$ distributions integrated over all decay angles. The
agreement is quite good; however, we are looking for more than just global 
agreement. Figure~\ref{fig:decay-angles-fit} shows the extracted angular 
distributions for the signal and background. The agreement with the 
generated distributions is excellent (see Fig.~\ref{fig:decay-angles-gen}).
A two-dimensional $\chi^2$ calculation comparing the generated and fit signal
histograms yields $\chi^2/${\em ndf} = 0.65 (where 
{\em ndf} is the degrees of freedom).
This comparison may not be ideal due to the relatively 
large errors which exist on the small bin occupancies; however, it is 
sufficient to demonstrate the quality of the signal-background separation. 
We can also compare the $Q$-factors extracted by the fits to the theoretical 
distributions from which our data was generated. 
Figure~\ref{fig:q-compare} shows that the extracted values are in very good
agreement with the generated ones.

We conclude this section by discussing the importance of quality control in
the fits. For this example, we performed 20,000 independent fits to extract the
$Q$-factors. To avoid problems which can arise due to fits 
not converging or finding local minima, each unbinned maximum likelihood fit
was run with three different sets of starting values for the parameters
$\vec{\alpha}$: 
(1)~100\% signal; 
(2)~100\% background; 
(3)~50\% signal, 50\% background.
In all cases, the fit with the best likelihood was used.
The $n_c$ events were then binned and a $\chi^2/${\em ndf}
was obtained. 
In about 2\% of the fits the $\chi^2/${\em ndf} was very large, 
a clear indicator that the
fit had not found the best estimators $\hat{\alpha}$.
For these cases, a binned $\chi^2$ fit was run to obtain the $Q$-factor. 

\subsection{Examining the Errors}

As discussed in Section~\ref{section:errors}, the covariance matrix obtained
from each event's fit can be used to obtain the uncertainty in $Q$, 
$\sigma_Q$, using Eq.~(\ref{eq:sigma-q}). The nature of our procedure leads to 
a high degree of correlation between neighboring events' $Q$-factors. This 
means that adding the uncertainties in quadrature would definitely 
underestimate the true error if the data is binned.
In Section~\ref{section:errors}, we showed how to properly handle these 
uncertainties and also argued that a decent approximation could be obtained by 
assuming 100\% correlation which provides an overestimate of the true error.

\begin{figure*}[t]
  \centering
  \subfigure[]{
    \includegraphics[width=0.49\textwidth]{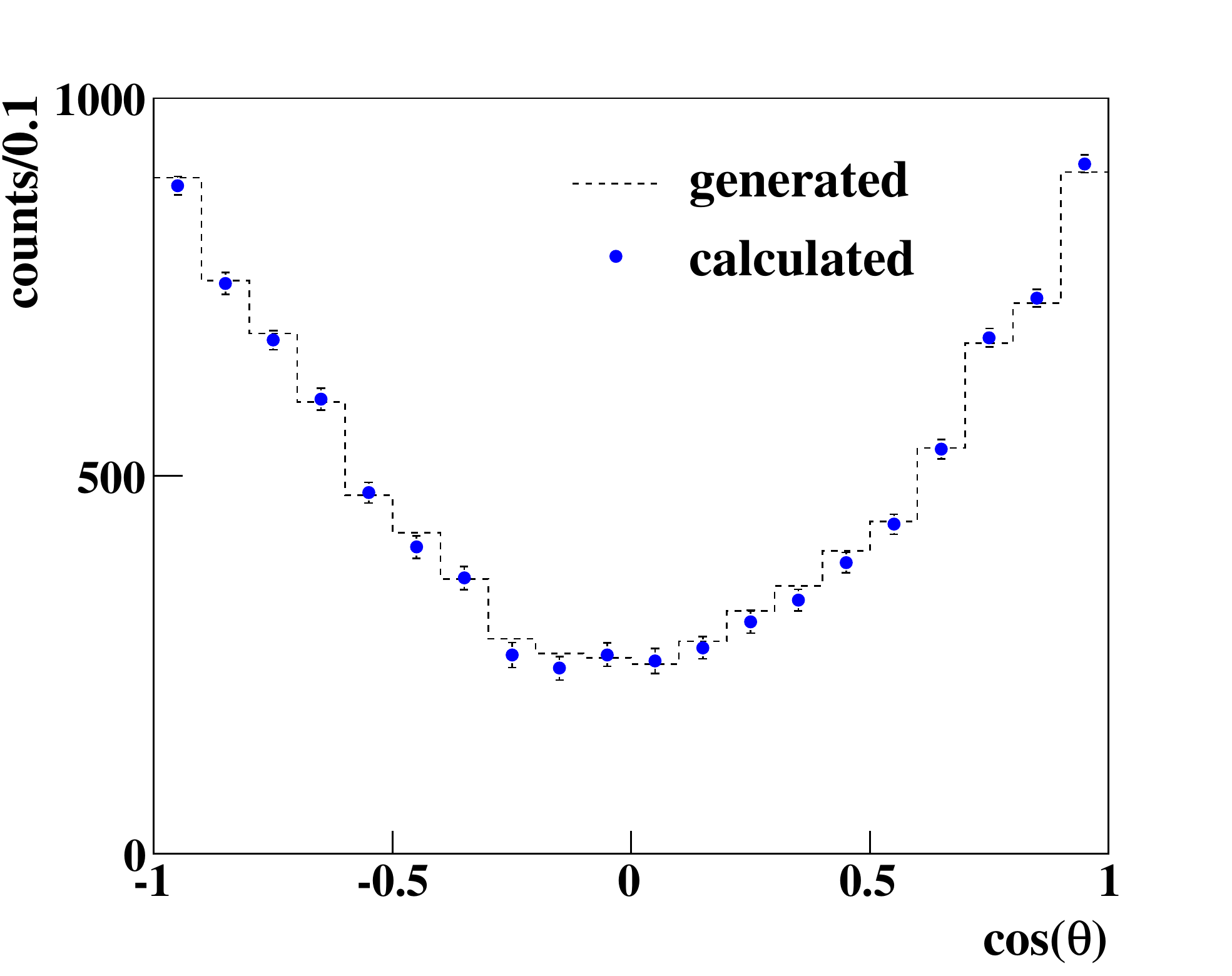}
  }
  \hspace{-0.02\textwidth}
  \subfigure[]{
    \includegraphics[width=0.49\textwidth]{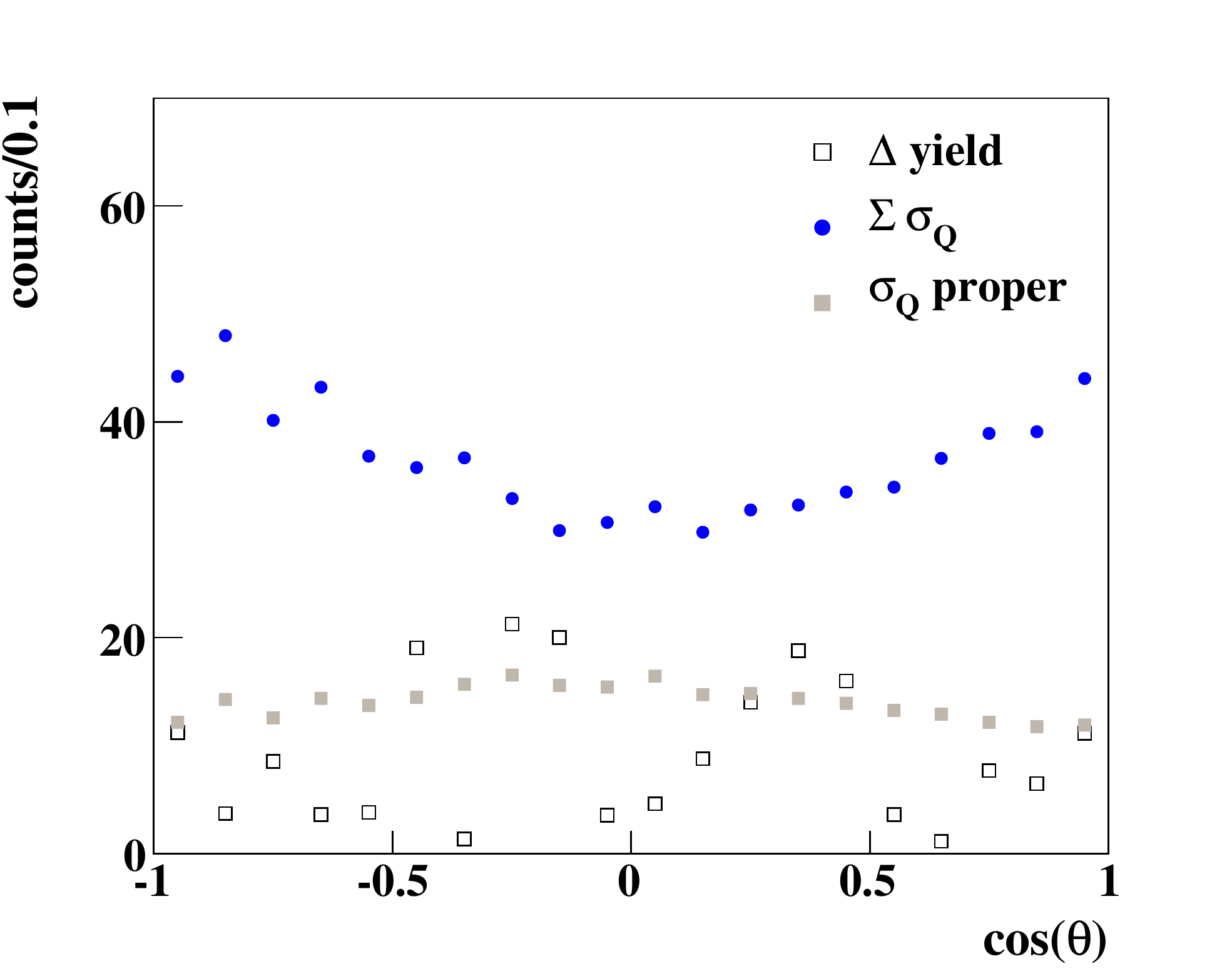}
  }
  \caption[]{\label{fig:compare-q-err}
    (Color Online)
    (a) Signal yield vs $\cos{\theta}$ for generated signal events (dashed) and
    all generated events weighted by $Q$-factors (blue circles). The error bars
    on the extracted yields were obtained using (\ref{eq:q-err-prop}).
    (b) Comparison of the true error on the signal yield, 
    $\Delta \mathcal{Y} = |\mathcal{Y}_{gen} - \mathcal{Y}_{calc}|$
    (open black squares), to the
    error bars obtained using Eq.~(\ref{eq:q-err-prop}) (gray squares) and 
    Eq.~(\ref{eq:q-err}) (blue circles).
  }
\end{figure*}

To examine the error bars in our toy example, we chose to project our data into
a one-dimensional distribution in $\cos{\theta}$. This was done to avoid 
bin occupancy issues which arise in the two-dimensional case due to limitations
in statistics.
Figure~\ref{fig:compare-q-err}(a) shows the comparison between the generated
and calculated $\cos{\theta}$ distributions. The agreement is excellent.
The error bars on the calculated
points were obtained using Eq.~(\ref{eq:q-err-prop}). 
For this study, we ignore 
the Poisson statistical uncertainty in the yield due to the fact that the 
number of generated events is known. In a real world analysis, these should be
included in the quoted error bars.

We can examine the quality of the error estimation by examining the difference
between the generated and calculated yields in each bin, $\Delta \mathcal{Y}$.
Figure~\ref{fig:compare-q-err}(b) shows the comparison between 
$\Delta \mathcal{Y}$, $\sigma_{\mathcal{Y}}$ obtained using 
Eq.~(\ref{eq:q-err-prop}) and $\sigma_{\mathcal{Y}}$ obtained assuming 100\% 
correlation.
As expected, the 100\% correlated errors provide an overestimate of 
$\Delta \mathcal{Y}$ in every bin, while the errors obtained using 
Eq.~(\ref{eq:q-err-prop}) provide an accurate calculation of the uncertainties.


\subsection{Extracting Observables}
\label{section:example:rho}

\begin{figure*}
  \centering
  \subfigure[]{
    \includegraphics[width=1.0\textwidth]{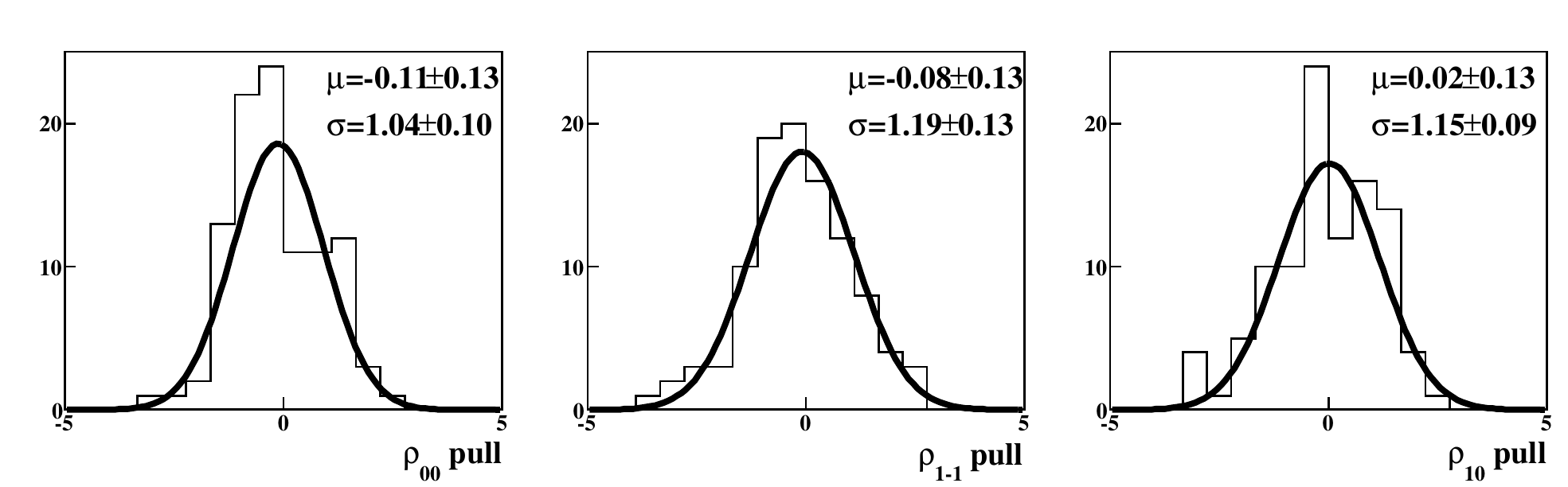}
  }
  \caption[]{\label{fig:pulls}
    Pull distributions for the spin density matrix elements extracted from an
    ensemble of datasets generated from the signal and background PDF's 
    given in Section~\ref{section:example}.
    The means and widths of the distributions are consistent with the 
    expected values (0 and 1, respectively).
  }
\end{figure*}

The goal of our model analysis is to extract the spin density matrix elements.
Binning the data would be undesirable due to limitations in statistics. 
Following Section~\ref{section:obs}, we can instead perform an event-based
maximum likelihood fit employing the $Q$-factors to handle the presence of
background in our data set.  The log likelihood is obtained from
Eq.~\ref{eq:log-L-Q-def} as
\begin{equation}
  \label{eq:log-L-Q}
  -\ln{\mathcal{L}} = -\sum\limits_i^{n} Q_i \ln{W(\theta_i,\phi_i)},
\end{equation}
where the sum is over all events (which contains 
signal and background).
Using the $Q$-factors obtained in Section~\ref{section:example:apply}, 
minimizing Eq.~(\ref{eq:log-L-Q}) yields
\begin{subequations}
  \begin{equation}
  \rho^0_{00} = 0.659\pm0.011 
  \end{equation}
  \begin{equation}
  \rho^0_{1-1} = 0.044\pm0.008 
  \end{equation}
  \begin{equation}
  Re\rho^0_{10} = 0.108\pm0.007,
  \end{equation}
\end{subequations}
where the uncertainties are purely statistical (obtained from the fit 
covariance matrix). The values extracted for the spin density matrix
elements are in excellent agreement with the values used to generate the
data given in Eq.~(\ref{eq:rho-gen}). To verify the accuracy of the 
statistical uncertainties, an ensemble of 100 example datasets was generated
and the analysis procedure repeated on each independently.  
The pull distributions obtained for the spin density matrix elements are shown 
in Fig.~\ref{fig:pulls}. The means and widths are consistent with the expected 
values.  Thus, no bias is found in the extracted observables and the 
statistical uncertainties obtained from the fits are correct.

\subsection{Choosing a Metric and a Value for $n_c$}
\label{section:example:nc}

\begin{figure}
  \begin{center}
  \subfigure[]{
    \includegraphics[width=0.49\textwidth]{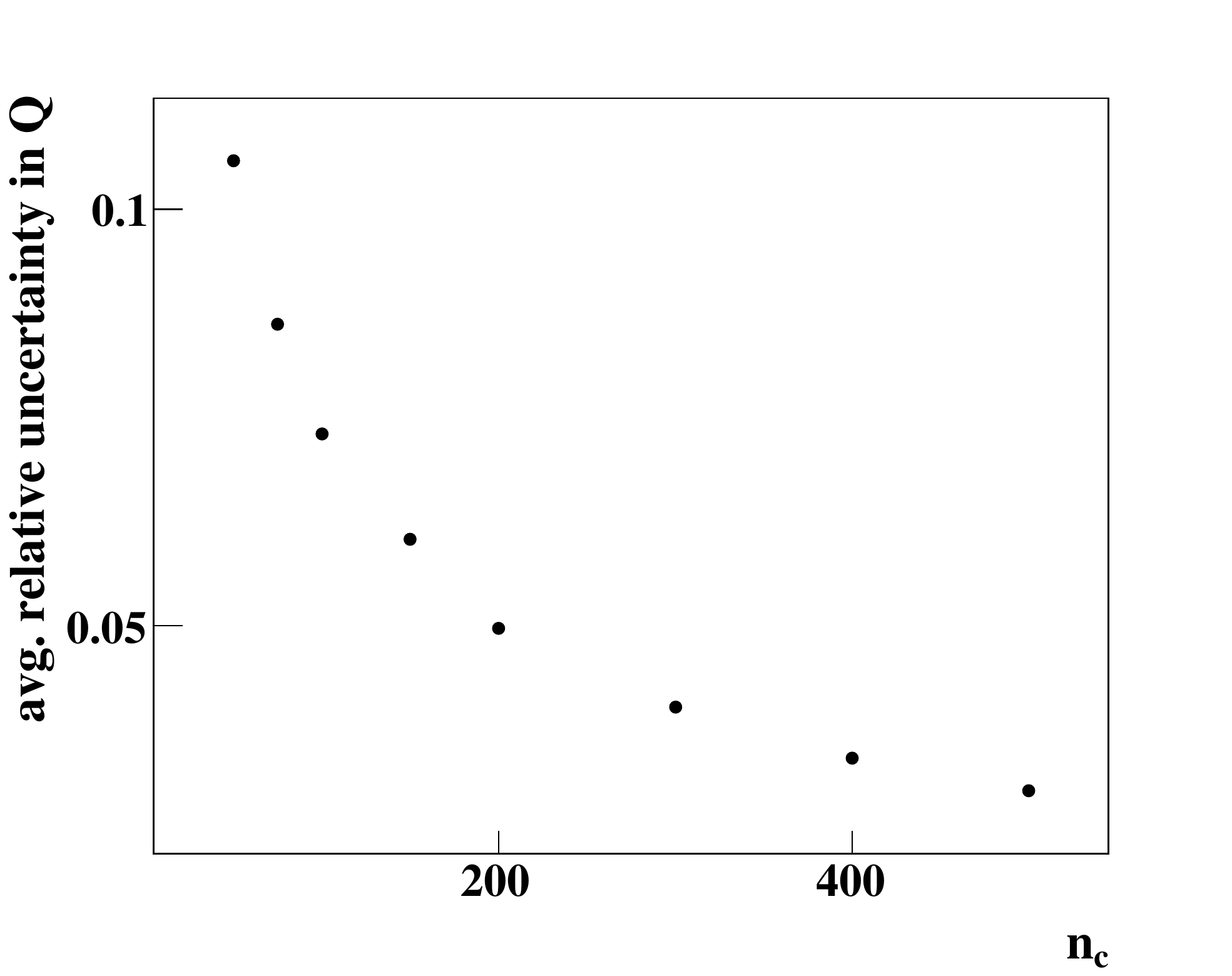}
  }
  \hspace{-0.02\textwidth}
  \subfigure[]{
    \includegraphics[width=0.49\textwidth]{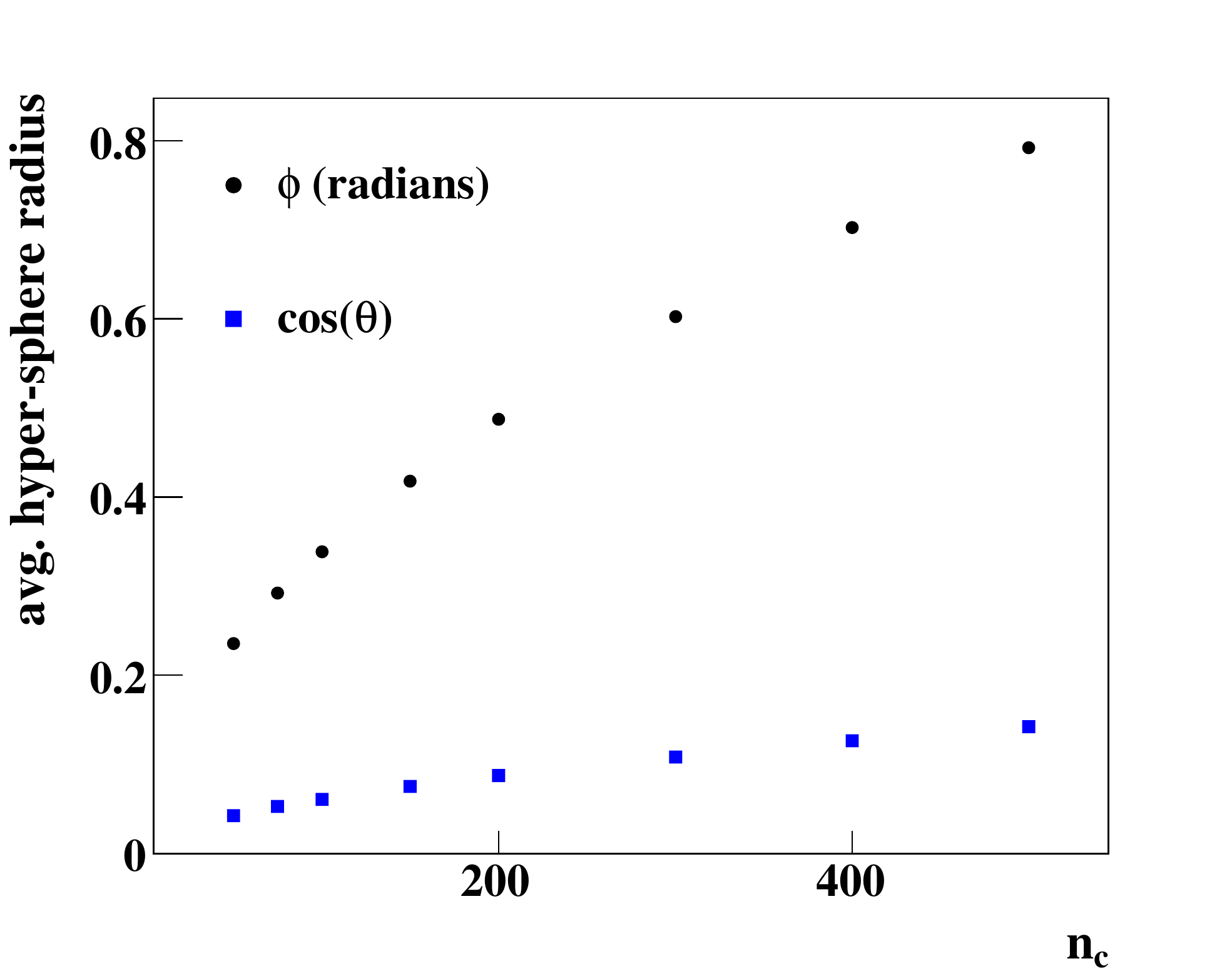}
  }
  \caption[]{\label{fig:nc}
    (a) Average relative uncertainty in $Q$ {\em vs} $n_c$. 
    (b) Average hyper-sphere radius {\em vs} $n_c$.
  }
  \end{center}
\end{figure}

In this section, we will examine the choices of the metric and of the value 
of $n_c$ 
(the number of nearest neighbor events used to determine the $Q$-factors).
As stated above, it is not possible to provide general prescriptions for the 
metric and the value of $n_c$ that will guarantee the validity of the method 
for every analysis.
Instead, these choices must be studied using the data and Monte Carlo data.
Figure~\ref{fig:nc}(a) shows the average relative uncertainty in $Q$ for 
different choices of $n_c$. 
As expected, this uncertainty increases as $n_c$ decreases.  
Keeping this uncertainty as small as possible dictates choosing a large value 
for $n_c$.  
If, however,
$n_c$ is large relative to $n$, then the method will average over large 
fractions of phase space. This can result in losses of the finer structure in 
the distributions.
Thus, there are two competing factors which should be considered when choosing
the value of $n_c$. The ratio of $n_c/n$ must be small enough to permit a true
extraction of the finer structure; however, the value of 
$n_c$ must be large enough such that the relative uncertainties in $Q$
are not too large. These constraints are analogous to those used to
choose a bin width in a binned analysis.

\begin{figure*}[t]
  \centering
  \subfigure[]{
    \includegraphics[width=0.49\textwidth]{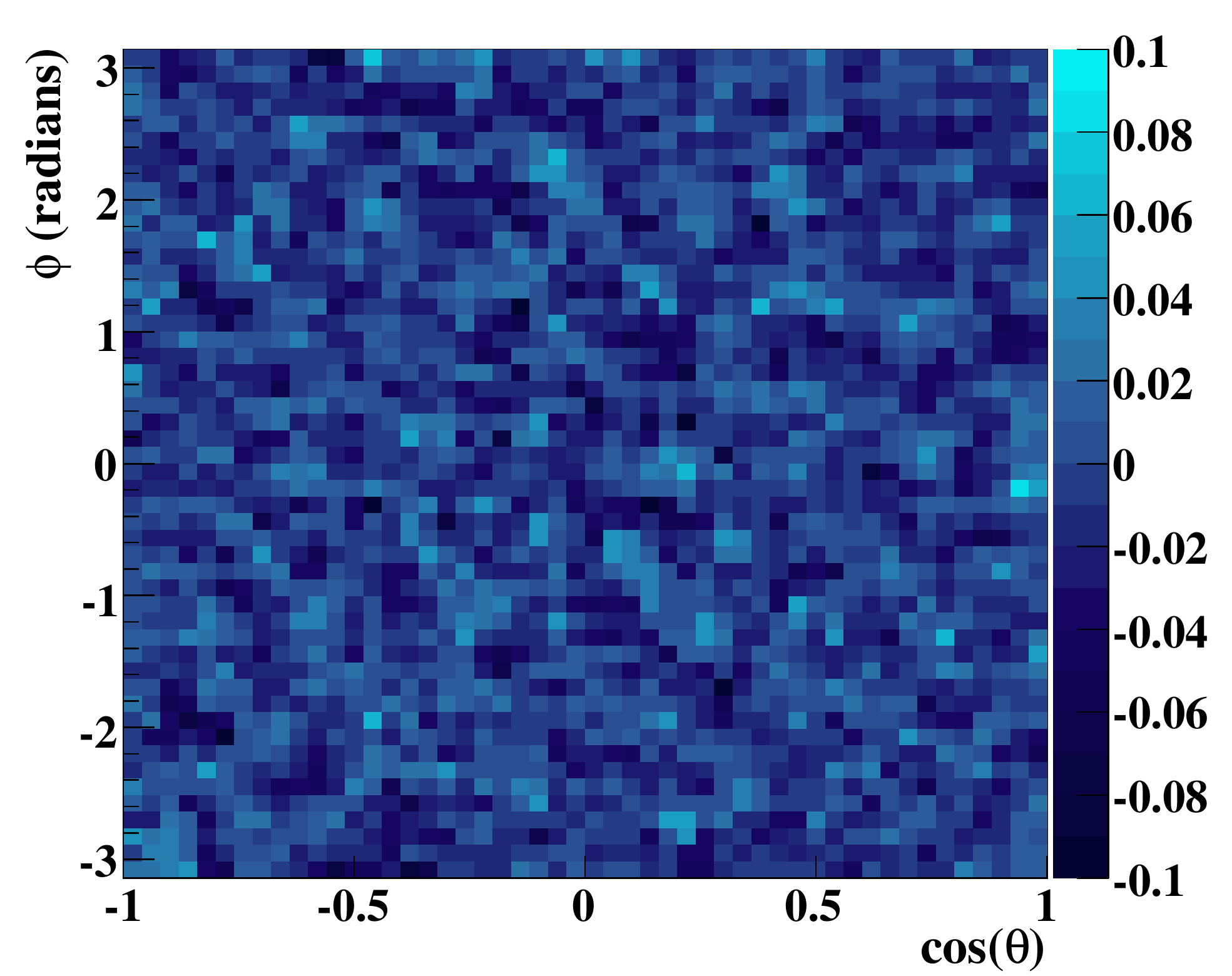}
  }
  \hspace{-0.02\textwidth}
  \subfigure[]{
    \includegraphics[width=0.49\textwidth]{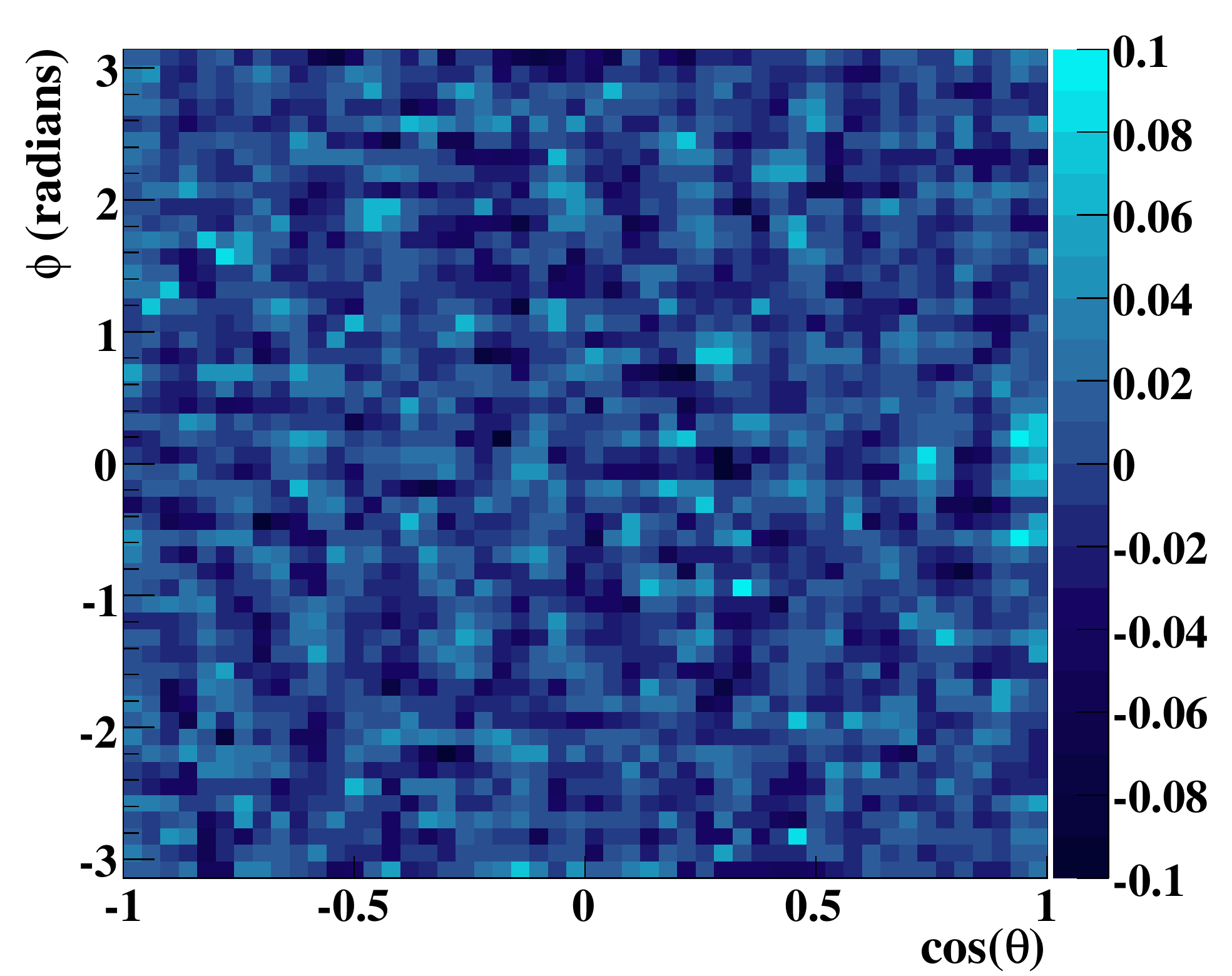}
  }
  \caption[]{\label{fig:compare-q-2d}
    (Color Online)
    (a) $\Delta Q$ vs. $\phi$ (radians) vs. $\cos{\theta}$: Mean 
    difference between the $Q$-factors extracted using $n_c=100$ and 
    $n_c=200$.  
    (b) $\Delta Q$ vs. $\phi$ (radians) vs. $\cos{\theta}$: Mean 
    difference between the $Q$-factors extracted using RMS's and
    {\em ranges} (see text for details) in the metric.
    In both cases, the differences in the $Q$-factors are consistent with 
    their statistical uncertainties, {\em i.e.} there is no sensitivity in the
    reconstructed signal distributions to these choices of $n_c$ or metric.
  }
\end{figure*}

Figure~\ref{fig:nc}(a) shows that statistical fluctuations in the $Q$-factors
become large (relative to variations in the physics) for $n_c < 100$.  
Figure~\ref{fig:nc}(b) shows the average radius (in $\phi$ and $\cos{\theta}$)
of the hyper-spheres used to collect the $n_c$ events for different choices of
$n_c$. From Fig.~\ref{fig:decay-angles-fit} we can estimate the size of the 
regions of phase space which are small enough such that the finer structure of 
the distributions is not lost. We then conclude that the choice $n_c \leq 200$
results in sufficiently small hyper-spheres.
Following the {\em ad-hoc} prescription discussed above, we combine these two
constraints which results in $100 \leq n_c \leq 200$.
Figure~\ref{fig:compare-q-2d}(a) shows a comparison of the $Q$-factors obtained
using $n_c=100$ and $n_c=200$. 
There is no apparent structure and the fluctuations are
consistent with the uncertainties on the $Q$-factors. 
Thus, the signal distributions reconstructed from these two 
choices of $n_c$ are statistically consistent. 
The spin density matrix elements extracted using the $Q$-factors obtained with 
$n_c=200$ are listed in Table~\ref{table:fit_rho_vals}.
The values are very close to those obtained using
$n_c=100$ in Section~\ref{section:example:rho} and well within the statistical
uncertainties. 

We now want to examine the effects of choosing a different metric. In the
previous sections, the RMS's of the variables were used in 
Eq.~(\ref{eq:dist}) to determine the distance between events.  
Another possible choice would be to simply use the {\em range} of the 
variables, {\em i.e.} replace $\sigma_{\cos{\theta}}$ by 2 and 
$\sigma_{\phi}$ by $2\pi$. 
Figure~\ref{fig:compare-q-2d}(b) shows a comparison of the $Q$-factors obtained
using RMS's in the metric and those obtained using ranges. 
We again conclude that the signal distributions reconstructed from these two 
choices are statistically consistent. 
The spin density matrix elements extracted using the $Q$-factors obtained with 
this alternative metric are listed in Table~\ref{table:fit_rho_vals}.
The results are again very close to those obtained in 
Section~\ref{section:example:rho}.

For this example, there is no physics-based motivation for adding dependence on
the coordinates to the metric; however, to further demonstrate the 
robustness of the method, we can replace the RMS's in Eq.~(\ref{eq:dist}) 
by the folloing quantities:
\begin{equation}
  \label{eq:var_metric}
  \sigma^2_{\phi} \rightarrow \left(\frac{2\pi^3}{3}\right)
  \left(1 - \frac{3\phi^2}{4\pi^2}\right), \qquad
  \sigma^2_{\cos{\theta}}\rightarrow \left(\frac{2}{3}\right)
  \left(1 - \frac{3}{4} \cos^2{\theta}\right).
\end{equation}
The spin density matrix elements extracted using the $Q$-factors obtained with 
this alternative {\em variable} metric are listed in 
Table~\ref{table:fit_rho_vals}.
The results are again very close to those obtained in 
Section~\ref{section:example:rho}.
Therefore, the extracted observables are stable provided the choices of metric 
and $n_c$ obey the {\em ad-hoc} constraints discussed above.

\begin{table}[h!]
\begin{center}
\begin{tabular}{c|cccc} 
  {} & generated & signal & $n_c=100$ [``RMS'' : ''range'' : ''var''] & $n_c=200$[``RMS'']
  \\\hline
  $\rho^0_{00}$ & 0.65 & 0.649 & [0.659 : 0.656 : 0.657] & 0.656 \\ 
  $\rho^0_{1-1}$ & 0.05 & 0.043 & [0.044 : 0.044 : 0.043] & 0.042 \\ 
  $Re\rho^0_{10}$ & 0.10 & 0.103 & [0.108 : 0.108 : 0.107] & 0.107 \\
  \hline
\end{tabular}
\\
\vspace{0.01\textheight}
\caption[]{\label{table:fit_rho_vals}
  $\rho^0_{MM'}$ values (fit and generated): 
  The signal results were obtained by fitting only the generated signal 
  events (unweighted).
  ``RMS'' refers to the metric constructed using the RMS's of the variables in 
  Eq.~(\ref{eq:dist}), ``range'' refers to simply using the ranges of the 
  variables, and ``var'' refers to the variable metric constructed using
  Eq.~(\ref{eq:var_metric}) (see text for details).  
  The statistical uncertainties on the extracted 
  quantities are $\sigma_{00} = 0.011$, $\sigma_{1-1} = 0.008$ and 
  $\sigma_{10} = 0.007$ for each of the fit conditions.
}
\end{center}
\end{table}

\subsection{Extending the Example}
\label{section:example:full-pwa}

To extend this example to allow for the case where the data is not binned in
production angle, we would simply need to include $\cos{\theta^{\omega}_{CM}}$
or $t$ in the vector of relevant coordinates, $\vec{\xi}$. 
To perform a full partial wave analysis on the data, we would also need to 
include any additional kinematic variables which factor into the partial wave 
amplitudes, {\em e.g.} the distance from the edge of the $\pi^+\pi^-\pi^0$ 
Dalitz plot (typically included in the $\omega$ decay amplitude). We would then
construct the likelihood from the partial waves and minimize 
$-\ln{\mathcal{L}}$ using the $Q$-factors obtained by applying our procedure 
including the additional coordinates. An example of this can be found 
in~\cite{cite:williams-thesis}.
\section{\label{section:conclusions}Conclusions}

In this paper, we have presented a procedure for separating signals from 
non-interfering backgrounds by determining, on an event-by-event basis, a 
quality factor ($Q$-factor) that a given event 
originated from the signal
distribution. This procedure is a generalization of the side-band subtraction 
method to higher dimensions which does not require any binning of the data.
We have shown that the $Q$-factors can be used as event 
weights in subsequent analysis procedures to allow more direct access to the
true signal distribution. For example, the $Q$-factors can be used to weight 
the log likelihood in event-based unbinned maximum likelihood fits. This leads
to background subtraction which is carried out automatically during the fits.

The method presented in this paper is meant to be applied to analyses where 
the distributions of the signal 
and background are unknown.  All that is required is that each can be 
parametrized in terms of (at least) one coordinate which must not be 
correlated with the remaining set of coordinates. While it may be possible
to overcome this limitation ({\em i.e.} it may be possible to account for the 
correlations in the PDFs), this topic is not explored in this paper as it is
likely to be very problem specific.
No knowledge about the distributions of the signal or background in any other 
coordinates is necessary. 
 
The validity of the method depends on the choices of two {\em ad-hoc} 
quantities: the metric and the number of nearest-neighbor events used to 
define each hyper-sphere. 
It is not possible to provide general prescriptions for these quantities
that will guarantee the validity of the method for every analysis.
Instead, they must be studied using the data and Monte Carlo data for each 
analysis individually to confirm that the method produces valid results.
The correlation distance varies according to the non-reference coordinates
which makes proper handling of the errors an intricate task. Thus,
the accuracy of the uncertainties on physical observables extracted from the 
data should be verified by a Monte Carlo study as well.
Given these caveats, it is clear that performing a detailed study using Monte 
Carlo data is an important prerequisite to using this method for any analysis.


\acknowledgments
This work was supported by grants from the United States Department of Energy
No. DE-FG02-87ER40315 and 
the National Science Foundation No. 0653316 through the 
``Physics at the Information Frontier'' program.

\end{document}